\documentclass[%
 reprint,
 superscriptaddress,
 amsmath,amssymb,
 aps,
 prx,
]{revtex4-2}

\usepackage{graphicx}
\usepackage{dcolumn}
\usepackage{amsmath}
\usepackage{mathtools}
\usepackage{hyperref}
\usepackage[dvipsnames]{xcolor}
\hypersetup{
  colorlinks=true,
  hypertexnames=false,
  linktocpage,
  colorlinks=true, 
  urlcolor=magenta!90!black,    
  linkcolor=blue!60!black, 
  citecolor=black!60 
}
\usepackage{amsthm,bm}
\usepackage{tikz}
\usepackage[braket, qm]{qcircuit}
\usepackage[ruled, vlined, linesnumbered]{algorithm2e}
\usepackage[percent]{overpic}
\usepackage{xprintlen}
\usepackage{layouts}
\usepackage{comment}
\usepackage[capitalize]{cleveref}

\theoremstyle{definition}

\DeclareMathOperator{\Tr}{Tr}

\begin{document}
\title{Continuous-angle logical rotations in the Steane code}
\author{Eric Huang}
 \affiliation{
 Joint Center for Quantum Information and Computer Science, NIST/University of Maryland,
College Park, Maryland 20742, USA.
}
\author{Daiwei Zhu}
\affiliation{IonQ, Inc., College Park, Maryland 20740, USA.}
\author{Matteo Ippoliti}
\affiliation{Department of Physics, The University of Texas at Austin, Austin, TX 78712, USA.}
\author{Christopher Monroe}
\affiliation{Duke Quantum Center, Department of Electrical \& Computer Engineering
and Department of Physics, Duke University, Durham, NC 27701, USA.}
\affiliation{IonQ, Inc., College Park, Maryland 20740, USA.}
\author{Michael J. Gullans}
 \affiliation{
 Joint Center for Quantum Information and Computer Science, NIST/University of Maryland,
College Park, Maryland 20742, USA.
}
\affiliation{National Institute of Standards and Technology, Gaithersburg, MD 20899, USA.}

\begin{abstract}
  We experimentally demonstrate continuous-angle logical \(Z\) rotations in the
  \([[7,1,3]]\) Steane code on the IonQ Forte trapped-ion processor.
  A round of the protocol applies a transversal physical \(Z\) rotation by \(\theta\),
  followed by Steane syndrome extraction and decoding, which induces a
  syndrome-dependent logical \(Z\) rotation.
  We analytically derive the effect of dephasing noise on the logical rotation
  angle and logical dephasing rate.
  Using logical Ramsey interferometry, we observe coherent syndrome-dependent
  logical rotations from a single round of the protocol.
  We find that the logical channel reconstructed from process tomography is a
  noisy logical \(Z\) rotation well explained by a dephasing model.
  We further implement a two-round protocol applying physical rotations
  \(+\theta\) and \(-\theta\),
  and observe cancellation of the total logical angle with low logical
  dephasing for repeated trivial syndromes.
  This constitutes a proof-of-principle demonstration of continuously tunable
  non-Clifford logical gates by transversal rotations and standard error
  correction in a small quantum code.
\end{abstract}
\date{\today}
\maketitle

\section{Introduction}

Quantum error correction is necessary for fault-tolerant quantum computation.
Transversal logical gates are naturally fault-tolerant since they do not spread
errors within code blocks,
but the Eastin-Knill theorem forbids universality from transversal gates
alone~\cite{eastinRestrictionsTransversalEncoded2009}.
Non-transversal gates may be implemented by magic-state
injection~\cite{gottesmanDemonstratingViabilityUniversal1999}, which consumes
high-quality magic states such as $\ket{T}$ states produced by magic state
distillation~\cite{bravyiUniversalQuantumComputation2005,meierMagicstateDistillationFourqubit2012,bravyiMagicstateDistillationLow2012,websterReducingOverheadQuantum2015,gidneyEfficientMagicState2019,litinskiMagicStateDistillation2019,itogawaEvenMoreEfficient2024}
or
magic state cultivation~\cite{gidneyMagicStateCultivation2024,vakninEfficientMagicState2025,chenEfficientMagicState2025,sahayFoldtransversalSurfaceCode2025,claesCultivatingStatesSurface2025}.
In situations where $T$ gates are the bottleneck while logical Clifford gates
are easy, a conventional strategy is Clifford+$T$ synthesis. This prescribes
the sequence of Clifford and $T$ gates to $\epsilon$-approximate a logical
rotation, such as $R_Z(\phi)$ with an arbitrary rotation angle $\phi$, with an
optimal T-count of $O\left(\log(1/\epsilon)\right)$.

Recent theoretical proposals point to a more efficient implementation for
continuous-angle logical gates,
especially for small rotation angles relevant for quantum simulation
applications~\cite{choiFaultTolerantNonClifford2023,heHighfidelityInitializationLogical2025,gavrielTransversalInjectionDirect2023,akahoshiPartiallyFaultTolerantQuantum2024,toshioPracticalQuantumAdvantage2024,ismailTransversalSTARArchitecture2025,yoshiokaTransversalGatesProbabilistic2025,chengEmergentUnitaryDesigns2024,bravyiCorrectingCoherentErrors2018,huangRobustPhaseContinuous2026}.
For many odd-distance codes with even-weight stabilizer generators, a
transversal physical rotation followed by syndrome extraction and decoding can
induce a syndrome-conditioned logical unitary
rotation~\cite{chengEmergentUnitaryDesigns2024}.
The logical angle is random since the syndrome is random, but it is knowable
since it depends only on the applied physical angle and the measured syndrome%
~\cite{bravyiCorrectingCoherentErrors2018}.
This is even fault-tolerant against stochastic Pauli errors and measurement
errors in certain low-angle regimes.~\cite{huangRobustPhaseContinuous2026}.
In particular for the surface code subjected to dephasing,
the logical dephasing rate relative to the logical rotation angle is
on average exponentially suppressed in the code distance.
By repeating this fault-tolerant logical rotation protocol over multiple
rounds, and strategically choosing the physical rotation angles adaptively, one
may efficiently approach a target logical rotation in the low-angle regime.

Trapped-ion processors are a natural platform for testing this idea since they
support arbitrary-angle single-qubit control and all-to-all entangling gate
connectivity.
The $[[7,1,3]]$ Steane code is a minimal setting for such an experiment.
It is the smallest CSS code that corrects arbitrary single-qubit errors and admits
well-studied gadgets for transversal Clifford gates, flagged state preparation,
and fault-tolerant syndrome extraction.
It thus serves as a convenient platform to study the continuous-angle
logical rotation protocol by transversal physical gates and error correction.

\begin{figure*}
  \begin{center}
    \begin{overpic}[]{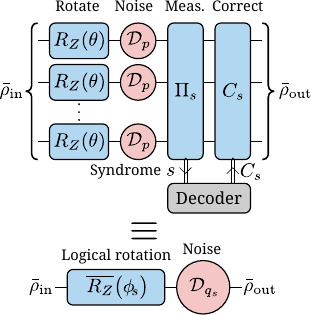}
      \put(0, 100){\textbf{(a)}}
    \end{overpic}
    \begin{overpic}[]{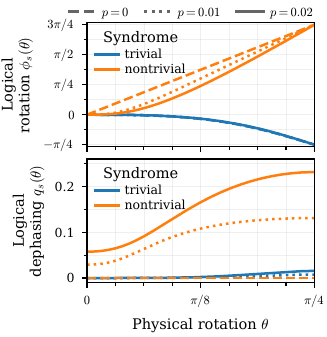}
      \put(0, 95){\textbf{(b)}}
    \end{overpic}
    \begin{overpic}[]{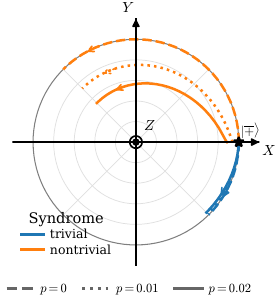}
      \put(0, 95){\textbf{(c)}}
    \end{overpic}
  \end{center}
  \caption{%
    Transversal logical rotation protocol in the Steane code with
    dephasing-only error model.
    \textbf{(a)}
    Single round of the continuous-angle rotation protocol takes a code state,
    applies a transversal rotation $R_Z(\theta)^{\otimes n}$
    with dephasing noise at rate $p$,
    measures a syndrome $s$ and
    applies the correction $C_s$ from the decoder.
    This is equivalent to a logical rotation by angle $\phi_s$ with
    dephasing at rate $q_s$, each of which depend only on the measured syndrome
    $s$,
    physical rotation angle $\theta$ and physical dephasing rate $p$.
    \textbf{(b)}
    Theoretical dependence of channel parameters as physical rotation angle
    $\theta$ for selected physical dephasing rates $p\in\{0, 0.01, 0.02\}$.
    Top: Logical rotation angle $\phi_s(p, \theta)$.
    Bottom: Logical dephasing rate $q_s(p, \theta)$.
    \textbf{(c)} The effect of the protocol is to rotate and contract
    the input code state $\ket{\overline{+}}$ (shown as a star)
    about the $Z$ axis of the logical Bloch sphere.
  }%
  \label{fig:schematic}
\end{figure*}

In this paper, we experimentally demonstrate encoded continuous-angle logical
$Z$ rotations in the $[[7,1,3]]$ Steane code on the IonQ Forte trapped-ion
processor using transversal physical rotations and Steane error correction.
Using logical Ramsey interferometry and logical process tomography, we observe
syndrome-conditioned logical rotations consistent with a simple theoretical
model.
We further implement a two-round protocol and observe approximate additivity of
the logical rotation angle with predictable noise accumulation.
These results provide a proof-of-principle demonstration of continuous-angle
logical control beyond the logical Clifford group in a small quantum code.
The protocol and its expected theoretical behavior are summarized in
\cref{fig:schematic}.

This paper is organized as follows.
In \cref{sec:background} we review the Steane code, flagged state preparation,
Steane syndrome extraction, the IonQ Forte hardware, and the transversal
continuous-angle rotation protocol.
In \cref{sec:theory} we derive the syndrome-conditioned logical channel under
coherent rotation and dephasing.
In \cref{sec:experiments} we present the one-round and two-round experimental
results.
We conclude with a discussion of the implications of this work in
\cref{sec:discussion}.

\section{Background}\label{sec:background}

\subsection{Steane code transversal gates}

The $[[7,1,3]]$ Steane code is the smallest CSS error-correcting code capable
of correcting arbitrary single-qubit errors.
It encodes $k=1$ logical qubit into $n=7$ physical qubits with distance $d=3$,
so that the minimum-weight logical operators have weight $3$.
The code borrows the parity-check matrix of the classical $[7,4,3]$
Hamming code,
\begin{align}
  H_X = H_Z =
  \begin{pmatrix}
  0 & 0 & 0 & 1 & 1 & 1 & 1\\
  0 & 1 & 1 & 0 & 0 & 1 & 1\\
  1 & 0 & 1 & 0 & 1 & 0 & 1
\end{pmatrix},\label{eqn:pcm}
\end{align}
to define the three $X$-type stabilizer generators
$X_4 X_5 X_6 X_7$, $X_2 X_3 X_6 X_7$ and $X_1 X_3 X_5 X_7$,
along with the three corresponding $Z$-type generators on the same supports.
A convenient choice of logical Pauli operators is $\overline{X}=X^{\otimes 7}$
and $\overline{Z}=Z^{\otimes 7}$.
Because the code is self-dual, it admits transversal implementations of
the Hadamard gate $\overline{H}=H^{\otimes 7}$,
as well as the $\overline{S}$ gate,
and CNOT, thereby realizing the
full logical Clifford group transversally.

\subsection{Code state preparation with flag qubit}

\begin{figure}[h]
    \input{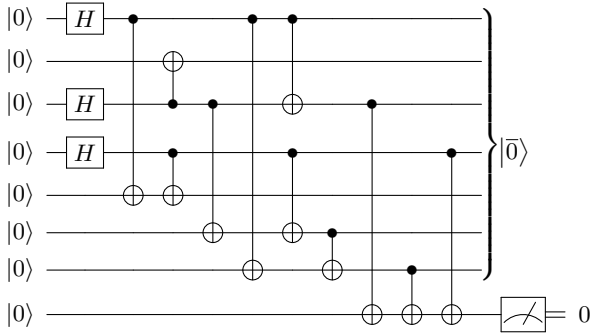}
    \caption{%
      A Steane code logical $\ket{\overline{0}}$ state may be prepared
      unitarily and made fault-tolerant by measuring the flag qubit and
      post-selecting on the flag qubit being measured in the $\ket{0}$
      state~\cite{zenQuantumCircuitDiscovery2025}.
    }%
    \label{fig:flag}
\end{figure}

The Steane code state $\ket{\overline{0}}$ may be unitarily prepared from
$\ket{0}^{\otimes n}$ by a non-fault-tolerant finite-depth circuit from its
stabilizer definitions.
An efficient technique to make this circuit fault tolerant
is to add just a single flag qubit,
which makes it capable of detecting a single error anywhere in the circuit.
The circuit in \cref{fig:flag} achieves this,
and with feedforward hardware capabilities,
the error may be corrected if the flag qubit is measured as $\ket{1}$.
However, for the experiments in this paper will only use its
error-detection properties for state preparation,
and postselect on untriggered flag qubits.

\subsection{Steane syndrome extraction}

\begin{figure}[h]
  \input{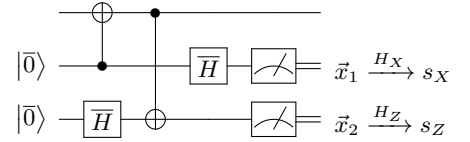}
  \caption{%
    Steane syndrome extraction uses transversal CNOT and Hadamard to
    fault-tolerantly measure the $X$-type and $Z$-type stabilizer generators of
    the Steane code using the classical parity check matrices on the measured
    ancilla outcomes.
    This assumes the ancilla $\ket{\overline{0}}$ states have been
    fault-tolerantly prepared.
  }%
  \label{fig:steanesyndrome}
\end{figure}

Due to their transversal CNOT gates,
CSS codes are amenable to Steane syndrome extraction and measurement,
which allows the fault-tolerant extraction of syndromes without rounds of
repeated measurements.
It is fault-tolerant against errors due to the intrinsic redundancy of the CSS
codewords and classical codes used to construct it.
The Steane syndrome extraction circuit is shown in \cref{fig:steanesyndrome}.
Using two ancilla block fault-tolerantly prepared in logical
$\ket{\overline{0}}$ states and transversal CNOT gates,
the $X$-syndrome and $Z$-syndromes may be measured non-destructively and
fault-tolerantly.
Upon measuring an outcome $\vec{x}_1$ on the first ancilla block,
the $X$-type syndrome is classically obtained by
$s_X = H_X \vec{x}_1$
where $H_X$ is the $X$-type parity check matrix.
Similarly, the other ancilla block gives the $Z$-type syndrome using the
$Z$-type parity check matrix $H_Z$.
The scheme can tolerate up to one fault anywhere in the circuit.

\subsection{IonQ Forte machine}

We perform the experiments in this work on the IonQ Forte trapped-ion quantum processor. IonQ Forte is a single-chain trapped-ion system supporting all-to-all connectivity, meaning any pair of qubits in the register can be directly entangled without SWAP routing. The Forte system uses the internal electronic states of each trapped ion as individual qubits and couples them together with shared motional modes as a data bus. Individual quantum gates are implemented via coherent Raman laser beams steered by acousto-optic deflectors~\cite{chenBenchmarkingTrappedionQuantum2024}.

A benchmarking study characterized IonQ Forte as a 30-qubit processor, and subsequent hardware updates extended the available register to 36 qubits while preserving the same connectivity model~\cite{chenBenchmarkingTrappedionQuantum2024}. For our specific experiment, the all-to-all connectivity eliminates the need for ancilla shuttling or long-range gate decomposition in the Steane code circuits. The 36-qubit register is sufficient to accommodate the seven data qubits, flag qubits for fault-tolerant state preparation, and the two seven-qubit ancilla blocks required for Steane syndrome extraction, with enough remaining qubits to implement the logical-basis rotations used in process tomography. These resource requirements make the Forte system a natural testbed for the transversal rotation protocol studied here.

\subsection{Transversal non-Clifford gate protocol}

\begin{figure}[h]
  \input{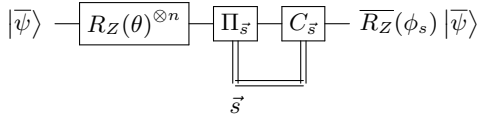}
  \caption{%
    Single round of the transversal rotation protocol.
    An encoded input state \(\ket{\overline{\psi}}\) is acted on by a transversal
    physical \(Z\) rotation \({R_Z(\theta)}^{\otimes n}\), followed by syndrome
    measurement with outcome \(\vec{s}\) and the corresponding decoder correction
    \(C_{\vec{s}}\).
    Conditioned on the measured syndrome, the net action on the code space is a
    logical rotation \(\overline{R_Z}(\phi_s)\), where the logical angle
    \(\phi_s(\theta)\) depends only on the syndrome $s$ and physical rotation
    angle $\theta$.
    With low noise, the logical operation will be a general channel.
  }%
  \label{fig:1rschema}
\end{figure}

Transversal rotations and error correction can implement fault-tolerant logical
rotations in certain regimes~\cite{huangRobustPhaseContinuous2026}.
The single-round rotate and correct protocol depicted in \cref{fig:1rschema} is
the building block which we adopt for this experiment.
Let $R_Z(\theta) := \exp(-i\theta Z/2)$ denote unitary $Z$ rotation by $\theta$.
Starting from an encoded Steane-code state,
we apply a transversal \(Z\) rotation $R_Z(\theta)^{\otimes n}$,
followed by Steane syndrome extraction and decoding.
Conditioned on measuring syndrome \(s\),
the decoder prescribes a Pauli correction \(C_s\),
returning the state to the code space.
If a CSS code has even-weight stabilizer generators,
odd $X$- and $Z$-distances, and the applied physical unitary commutes with the
time-reversal transformation
$T = \mathcal{K}\prod_j (iY_j)$
where $\mathcal{K}$ denotes complex conjugation~\cite{chengEmergentUnitaryDesigns2024},
then the resulting logical channel conditioned on the syndrome will be unitary.

Under physical single-qubit rotations, which are invariant under time-reversal,
the Steane code satisfies all of these requirements.
Its \(X\)- and \(Z\)-type stabilizer generators all have weight \(4\),
and it has odd distances
$d_x=d_z=3$.
For the coherent rotations considered in this work,
we restrict ourselves to $Z$ rotations,
so the logical channel is a unitary rotation $\overline{R_Z}(\phi_s(\theta))$.

Although the logical rotation angle $\phi_s(\theta)$ is random,
it is knowable since it depends only on the physical rotation
angle $\theta$ and on the observed syndrome $s$.
By iterating this protocol multiple times and strategically choosing $\theta$
each time,
we may target the desired logical rotation angle with minimal cost.

In the multi-round protocol, one applies a sequence of transversal rotations
with angles \(\theta_1,\theta_2,\ldots\),
interleaved with syndrome extraction and corrections after each round.
The logical angle accumulated after several rounds is then the sum
of the syndrome-conditioned contributions from each round.
More generally, the adaptive protocol in \cite{huangRobustPhaseContinuous2026}
chooses each subsequent physical rotation angle using the previously observed
syndromes, thereby steering the encoded qubit toward a desired target logical
rotation while avoiding the exponential overhead of simple postselection.
Keeping track of the accumulated noise, an optimized multi-round protocol with
reset provides a viable path to continuously tunable logical non-Clifford
rotations using only transversal coherent rotations, syndrome measurements, and
classical feedforward.

In the present experiment we implement the simplest version of this
idea on the \( [[7,1,3]] \) Steane code.
Because the code distance is small, the protocol should be viewed as a
proof-of-principle demonstration rather than a scalable fault-tolerant
realization.
In addition, the IonQ Forte experiments reported here do not use real-time
decoder feedforward within the circuit.
Instead, syndrome information is utilized in classical postprocessing to
identify the corresponding logical outcome.
Our one-round and two-round demonstrations therefore probe the syndrome-resolved
logical channels generated by the protocol, while leaving fully adaptive
feedforward implementations to future hardware with lower-latency classical
control.

\section{Ideal and Theoretical Logical Channel}\label{sec:theory}

In this section we derive the logical channel of the protocol under ideal
and noisy conditions assuming a simple dephasing-only noise model as depicted
in~\cref{fig:schematic}.

Suppose a CSS code state is subjected to coherent rotation $R_Z(\theta)$,
dephasing noise at rate $p$, syndrome extraction to obtain a syndrome $s$,
a Pauli correction $C_s$.
We model the physical single-qubit channel as a composition of a coherent
rotation $R_Z(\theta)$ by angle \(\theta\) and dephasing channel
$\mathcal{D}_p$ at rate \(p\),
\begin{align}
  \mathcal{N}_{p,\theta}(\rho)
  &=
  R_Z(\theta)\left[(1-p)\rho + p Z\rho Z\right]R_Z(\theta)^\dagger.
\end{align}
In the ideal case of~\cref{fig:1rschema} we would have $p=0$,
but we will first derive the general case.

For an encoded logical state \(\rho\) in the code space, the
subnormalized branch conditioned on obtaining syndrome \(s\) is
\begin{align}
  \mathcal{E}_s(\rho)
  :=
  C_s \Pi_s \mathcal{N}_{p,\theta}^{\otimes n}(\rho)\Pi_s C_s^\dagger,
\end{align}
where \(\Pi_s\) projects onto the subspace of syndrome \(s\) and \(C_s\) is the
corresponding decoder correction. Restricted to the logical code space, this
branch takes the form
\begin{align}
  \mathcal{E}_s(\rho)
  &=
  p_s(p,\theta)\,\mathcal{N}_{q_s,\phi_s}(\rho),
\end{align}
where \(p_s(p,\theta)\) is the probability of obtaining syndrome \(s\), and
\begin{align}
  \mathcal{N}_{q_s,\phi_s}(\rho)
  &=
  R_Z(\phi_s)\left[(1-q_s)\rho + q_s Z\rho Z\right]R_Z(\phi_s)^\dagger
\end{align}
is the normalized logical channel with logical rotation
\(\phi_s(p,\theta)\) and logical dephasing rate \(q_s(p,\theta)\).
The syndrome probability is obtained from the trace of the subnormalized branch,
\begin{align}
  p_s(p,\theta)=\Tr\left[\mathcal E_s(\rho)\right],
\end{align}
for any normalized logical input state \(\rho\),
such as \(\ket{\overline{0}}\bra{\overline{0}}\).
It is convenient to define the syndrome-conditioned logical coherence factor
\begin{align}
  \eta_s(p, \theta) &= 
  \bra{\overline{0}}\mathcal{E}_s\left(|\overline{0}\rangle\langle \overline{1}|\right)\ket{\overline{1}}\\
  &=
  p_s(p,\theta)\left(1-2q_s(p,\theta)\right)e^{-i\phi_s(p,\theta)}.
\end{align}
Knowing these complex factors is enough to determine the logical channel
parameters via
\begin{align}
  \phi_s(p,\theta) &= -\arg \eta_s(p,\theta),\label{eqn:phis} \\
  q_s(p,\theta) &=
  \frac{1}{2}\left(1-\frac{|\eta_s(p,\theta)|}{p_s(p,\theta)}\right)\label{eqn:qs},
\end{align}
with \(\phi_s=0\) and \(q_s=0\) by convention whenever \(p_s=0\).

One may compute $\eta_s$ by expanding $\ket{\overline{0}}\bra{\overline{1}}$ in the
computational basis as a linear combination of $\ket{x}\bra{y}$ bit string
basis elements.
The physical channel acts
independently on each qubit so dephasing contributes a factor of
\(\lambda:= 1- 2p\)
wherever \(x_j\neq y_j\), giving the overall factor
\(\lambda^{|x\oplus y|}\), while the transversal rotation contributes the phase
\(e^{i\theta(|x|-|y|)}\). Thus the physical channel acts diagonally like
\begin{align}
  \mathcal{N}_{p,\theta}^{\otimes n}(\ket{x}\bra{y})
  =
  \lambda^{|x\oplus y|}e^{i\theta(|x|-|y|)}\ket{x}\bra{y}.
\end{align}
As the correction \(C_s=\otimes_{j=1}^{n} Z^{z_j}\) for some
$z \in \mathbb{Z}_2^n$ is also diagonal, it contributes a sign like
\begin{align}
  C_s \ket{x}\bra{y} C_s^\dagger
  =
  (-1)^{z\cdot(x\oplus y)} \ket{x}\bra{y}.
\end{align}
The correction $C_s$ maps the syndrome-\(s\) subspace to the code space,
so \(C_s\Pi_s=\Pi_0 C_s\) where $\Pi_0$ is the projector onto the code space,
thus giving
\begin{align}
  \bra{\bar 0}\mathcal E_s(\ket{x}\bra{y})\ket{\bar 1}
  =
  \bra{\overline{0}}\Pi_0 C_s \mathcal{N}_{p,\theta}^{\otimes n}(\ket{x}\bra{y}) C_s^\dagger \Pi_0\ket{\overline{1}}\nonumber\\
  \qquad = \lambda^{|x\oplus y|}e^{i\theta(|x|-|y|)}
  (-1)^{z\cdot(x\oplus y)}
  \langle \bar 0|x\rangle \langle y|\bar 1\rangle,
\end{align}
where we have used
\(\bra{\bar 0}\Pi_0=\bra{\bar 0}\) and \(\Pi_0\ket{\bar 1}=\ket{\bar 1}\)
for code words $\ket{\overline{0}}$ and $\ket{\overline{1}}$.
Thus $\eta_s(p, \theta)$ is a linear combination of monomials
$\lambda^a e^{i\theta b}$ for integers $a\ge 0$ and $b$.
A similar calculation applies for the real syndrome probabilities
$p_s(p, \theta)$.
The coefficients may thus be obtained from the code words of the CSS code,
which are uniform superpositions of classical code words,
and the decoder corrections as bit strings.

\subsection{Steane code logical channel}

For the Steane code with the optimal decoder,
the coherence factors are
\begin{align}
  \eta_{\mathrm{t}}(p,\theta)
  &=
  \frac{e^{i\theta}}{64}
  \left[
    14\lambda^3\left(3+e^{-4i\theta}\right)
    +
    \lambda^7\left(7+e^{-8i\theta}\right)
  \right],
\end{align}
for the trivial syndrome, and
\begin{align}
  \eta_{\mathrm{n}}(p,\theta)
  &=
  \frac{e^{i\theta}}{64}
  \left[
    2\lambda^3\left(3+e^{-4i\theta}\right)
    -
    \lambda^7\left(7+e^{-8i\theta}\right)
  \right]
\end{align}
for any nontrivial syndrome. These are identical for all nontrivial syndromes
by symmetry.
The trivial-syndrome probability is
\begin{align}
  p_{\mathrm{t}}(p,\theta)
  &=
  \frac{1}{8}
  +
  \frac{7}{32}\lambda^4\left[3+\cos (4\theta)\right],
\end{align}
while the probability of any particular nontrivial syndrome is
\begin{align}
  p_{\mathrm{n}}(p,\theta)
  &=
  \frac{1-p_{\mathrm{t}}(p,\theta)}{7}.
\end{align}
The ensuing logical rotation angle $\phi_s(p,\theta)$ and logical dephasing
$q_s(p,\theta)$ by substituting the above into \cref{eqn:phis} and \cref{eqn:qs}
respectively are the curves plotted in~\cref{fig:schematic}(b).

\subsection{Ideal noiseless case}
In the ideal noiseless case \(p=0\) we have \(\lambda=1\),
so these coherence factors become
\begin{align}
  \eta_{\mathrm{t}}^{\mathrm{ideal}}(\theta)
  &=
  \frac{e^{i\theta}}{64}\left(7+e^{-4i\theta}\right)^2, \\
  \eta_{\mathrm{n}}^{\mathrm{ideal}}(\theta)
  &=
  -\frac{e^{i\theta}}{64}\left(1-e^{-4i\theta}\right)^2,
\end{align}
for the trivial syndrome and any nontrivial syndrome, respectively.
The corresponding ideal syndrome probabilities are
\begin{align}
  p_{\mathrm{t}}^{\mathrm{ideal}}(\theta)
  &=
  \frac{25+7\cos(4\theta)}{32}, \\
  p_{\mathrm{n}}^{\mathrm{ideal}}(\theta)
  &=
  \frac{1-\cos(4\theta)}{32}.
\end{align}
Because \(\eta_s^{\mathrm{ideal}}=p_s^{\mathrm{ideal}}e^{-i\phi_s^{\mathrm{ideal}}}\),
the ideal logical angles are
\begin{align}
  \phi_{\mathrm{t}}^{\mathrm{ideal}}(\theta)
  &=
  -\arg\!\left[e^{i\theta}\left(7+e^{-4i\theta}\right)^2\right], \\
  \phi_{\mathrm{n}}^{\mathrm{ideal}}(\theta)
  &=
  3\theta
\end{align}
modulo the chosen branch convention for the argument,
and the ideal logical dephasing rates vanish,
\begin{align}
  q_{\mathrm{t}}^{\mathrm{ideal}}(\theta)
  =
  q_{\mathrm{n}}^{\mathrm{ideal}}(\theta)
  =
  0.
\end{align}

These formulas are the ideal curves depicted throughout the paper as dashed
lines.
The ideal limit gives the target syndrome-conditioned logical rotations and
syndrome probabilities, while the noisy expressions quantify how physical
dephasing modifies both the logical coherence and the branch weights.
In particular, the nontrivial syndrome branches remain symmetry-equivalent,
whereas the trivial branch exhibits a distinct nonlinear dependence on
\(\theta\).

\section{Experiments}\label{sec:experiments}

\subsection{Ramsey interferometry}
\begin{figure}[h]
  \input{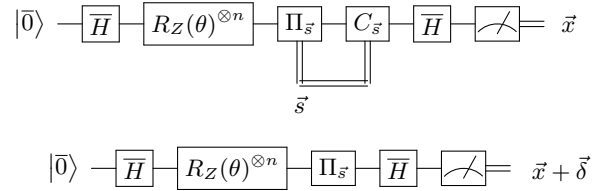}
  \caption{%
    \textbf{Top:} The logical Ramsey interferometry circuit prepares the logical
    $\ket{\overline{+}}$ state, performs one round of the $Z$-rotation protocol,
    and measures in the logical $\overline{X}$ basis.
    Steane syndrome extraction is used for $\Pi_{\vec{s}}$.
    \textbf{Bottom:} Equivalent circuit with the decoder correction
    $C_s=\otimes_j Z^{z_j}$ propagated to a flip in the classical outcome
    $\vec{\delta} = H_X \vec{z}$, which may be removed in postprocessing.
  }%
  \label{fig:protocol}
\end{figure}

Ramsey interferometry was originally introduced to measure phase accumulation in
two-level systems~\cite{ramseyMolecularBeamResonance1950}.
In our first experiment, we perform \emph{logical} Ramsey interferometry on the
one-round protocol to probe syndrome-conditioned coherent logical $Z$
rotations.
The principle is that an initial $\ket{\overline{+}}$ state acquires a phase in
the logical $XY$ plane, which is converted into oscillations of the logical
$\overline{X}$ measurement outcome.
The circuit in \cref{fig:protocol} outputs an \(n\)-bit string \(\vec{x}\),
which is classically decoded to obtain the logical measurement outcome.
The Steane code is self-dual, so
\(\ket{\overline{0}} \propto \sum_{x\in C}\ket{x}\), where \(C\) is the
\([7,4,3]\) Hamming code generated by the rows of the parity-check matrix
\(H_X\) in \cref{eqn:pcm}.
Thus, if \(\vec{x}\in C\), we assign logical outcome \(0\).
Likewise,
\(\ket{\overline{1}} \propto \sum_{x\in C}\ket{x\oplus 1111111}\),
so if \(\vec{x}\in C\oplus 1111111\), we assign logical outcome \(1\).
More generally, for arbitrary \(\vec{x}\), we decode to the logical bit of the
nearest of these two cosets.

Since the protocol involves feedforward error correction, it is convenient to
defer the correction to classical postprocessing, as shown in the bottom circuit of
\cref{fig:protocol}.

In our setting, we prepare the logical $\ket{\overline{+}}$ state, apply a
single round of the transversal rotation and error-correction protocol, and
then measure in the logical $\overline{X}$ basis.
Conditioned on the measured syndrome $s$, the ideal signal is
\begin{align}
  P(\overline{X}=+1 \mid s)
  =
  \frac{1+\cos\phi_s(\theta)}{2},
\end{align}
where the logical rotation angle $\phi_s(\theta)$ depends on the physical
rotation angle $\theta$ and the observed syndrome branch.

\begin{figure}[h]
  \begin{center}
    \includegraphics{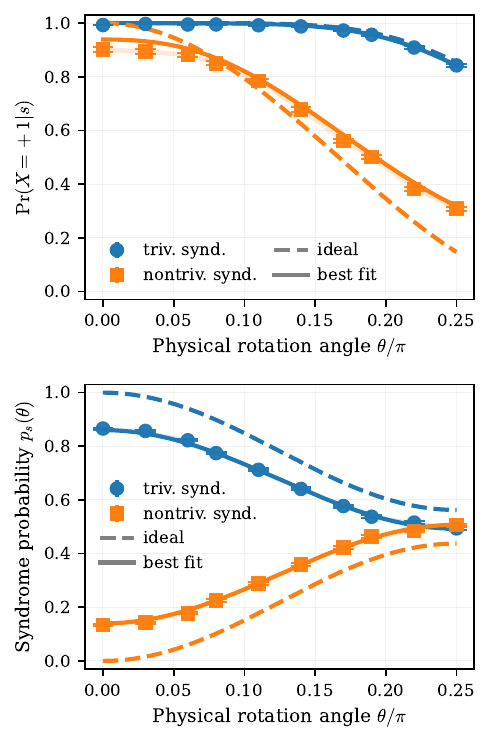}
  \end{center}
  \caption{
    Logical Ramsey experiment with input state \(\ket{\overline{+}}\) using the
    one-round protocol.
    \textbf{Top:}
    Syndrome-conditioned Ramsey signal \(P(\overline{X}=+1\mid s)\) versus physical
    rotation angle \(\theta\) for the trivial (blue) and aggregated nontrivial
    (orange) syndrome groups.
    Points with error bars show the data, dashed curves show the ideal
    prediction, and solid curves show the best-fit dephasing model.
    \textbf{Bottom:}
    Corresponding syndrome-group probabilities \(p_s(\theta)\) for the same two
    branches, with the same line conventions.
    The best-fit physical dephasing parameter is \(p=(2.13\pm0.07)\%\).
  }\label{fig:ramsey}
\end{figure}

The measured Ramsey fringes are shown in \cref{fig:ramsey}, where we also plot
the ideal noiseless curves as dashed lines.
The top panel shows the syndrome-conditioned probability
$P(\overline{X}=+1 \mid s)$ as a function of physical rotation angle $\theta$
for the trivial syndrome branch and the aggregate of all nontrivial syndrome
branches.
Both branches exhibit clear oscillations, demonstrating that a single round of
the protocol implements a coherent logical rotation whose angle is controlled by
the applied physical rotation.

The bottom panel of \cref{fig:ramsey} shows the corresponding syndrome-group
probabilities $p_s(\theta)$ for each branch.
As expected, the trivial syndrome is most likely for small rotation angles,
while the nontrivial syndromes become more common as the physical rotation angle
increases.

We fit the $p_s(\theta)$ data to the dephasing-only model introduced in
\cref{sec:theory}, obtaining a best-fit physical dephasing rate
$p = (2.13 \pm 0.07)\%$.
The resulting fit, shown as a solid line in \cref{fig:ramsey}, is in excellent
agreement with the observed syndrome probabilities.
When this same fitted value of $p$ is used to plot the Ramsey signal, it also
shows strong qualitative agreement with the measured fringes.
This provides an independent consistency check on the extracted parameter $p$
and suggests that the observed Ramsey data are well described by coherent
transversal rotation together with modest dephasing.

These results show that a single round of transversal physical $Z$ rotation
followed by Steane error correction produces a knowable,
syndrome-conditioned logical $Z$ rotation.
This motivates the fuller channel characterization presented next via logical
process tomography.

\subsection{Process tomography for single logical rotation}

\begin{figure}[h]
  \input{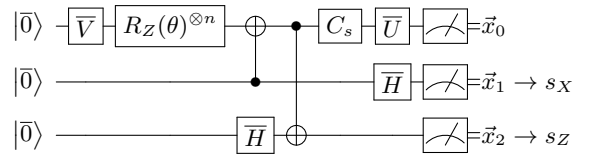}
  \caption{%
    One-round logical process tomography circuit.
    The logical Clifford \(\overline{V}\) prepares one of the four input states
    \(\ket{\overline{0}}, \ket{\overline{1}}, \ket{\overline{+}},
    \ket{\overline{+i}}\).
    After one round of the transversal rotation protocol with syndrome
    measurement \(\Pi_{\vec{s}}\) and decoder correction \(C_s\), a final
    logical Clifford \(\overline{U}\) selects measurement in the logical
    \(\overline{X}\), \(\overline{Y}\), or \(\overline{Z}\) basis.
    Repeating over all choices of \(\overline{V}\) and \(\overline{U}\)
    reconstructs the syndrome-conditioned logical channel \(\mathcal{E}_s\).
  }%
  \label{fig:1rcirc}
\end{figure}

Process tomography generalizes the Ramsey experiment by reconstructing the full
effective single-qubit logical channel conditioned on each syndrome branch.
Instead of using only the logical input state \(\ket{\overline{+}}\) and
measuring only in the logical \(\overline{X}\) basis, we prepare a
tomographically complete set of logical input states
\(\ket{\overline{0}}\), \(\ket{\overline{1}}\),
\(\ket{\overline{+}}\), and \(\ket{\overline{+i}}\), and measure the output in
the logical \(\overline{X}\), \(\overline{Y}\), and \(\overline{Z}\) bases.
This gives sufficient information to fully reconstruct the quantum channel
using standard
techniques~\cite{nielsenQuantumComputationQuantum2012,benentiSimpleRepresentationQuantum2009,mohseniQuantumprocessTomographyResource2008,cenedeseCorrectingCoherentErrors2023}.
For each physical rotation angle \(\theta\), we prepare the logical input
state, apply a single round of the protocol, condition on the measured syndrome
\(s\), and perform the final logical-basis measurement.
The full tomography circuit is shown in \cref{fig:1rcirc},
where to prepare the input state we apply the logical Clifford
\(\overline{V}\), chosen as \(I\) for \(\ket{\overline{0}}\),
\(\overline{X}\) for \(\ket{\overline{1}}\),
\(\overline{H}\) for \(\ket{\overline{+}}\), and
\(\overline{S}\,\overline{H}\) for \(\ket{\overline{+i}}\).
Similarly, we apply the logical Clifford \(\overline{U}\) before the final
measurement to choose the output basis, using \(\overline{H}\) for
\(\overline{X}\), \(\overline{H}\,\overline{S}^\dagger\) for
\(\overline{Y}\), and \(I\) for \(\overline{Z}\).
Together, these preparations and measurements determine the effective logical
channel \(\mathcal{E}_s\) for each syndrome branch.

\begin{figure}[h]
  \input{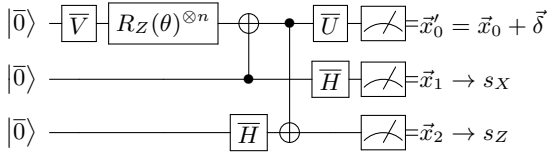}
  \caption{%
    Equivalent one-round logical process tomography circuit in which the
    syndrome-dependent decoder correction \(C_s\) is not applied physically, but
    is instead absorbed into the classical interpretation of the final
    measurement outcome.
    This avoids real-time feedforward while preserving the same
    syndrome-conditioned logical channel as in \cref{fig:1rcirc}.
  }
  \label{fig:1rcircpp}
\end{figure}

Ideally, after measuring the syndrome \(s\), one would apply the corresponding
Pauli correction \(C_s\) by feedforward before the final logical-basis
measurement.
To avoid real-time feedforward, we instead use the equivalent circuit of
\cref{fig:1rcircpp}, in which the correction is absorbed into the
interpretation of the final measurement record in classical postprocessing.

We can extract the logical rotation angle $\phi_s$ from the tomographically
reconstructed channel $\mathcal{E}_s$ by decomposing the reconstructed channel
into a coherent logical \(Z\) rotation and a residual noise channel,
\begin{align}
  \mathcal{E}_s = \mathcal{E}'_s \circ \mathcal{R}_Z(\phi_s),
\end{align}
where \(\phi_s\) is the best-fit logical rotation angle and
\(\mathcal{E}'_s\) captures the remaining incoherent part of the channel.
Numerically, \(\phi_s\) is chosen to minimize the Frobenius distance between
the affine Bloch matrix of \(\mathcal{E}'_s\) and the identity.
We use this convenient, stable least-squares criterion to separate the coherent
rotation; the residual channel is subsequently characterized by its average
gate infidelity, an operationally meaningful metric.

\begin{figure}[h]
  \begin{center}
    \begin{overpic}[width=0.95\columnwidth]{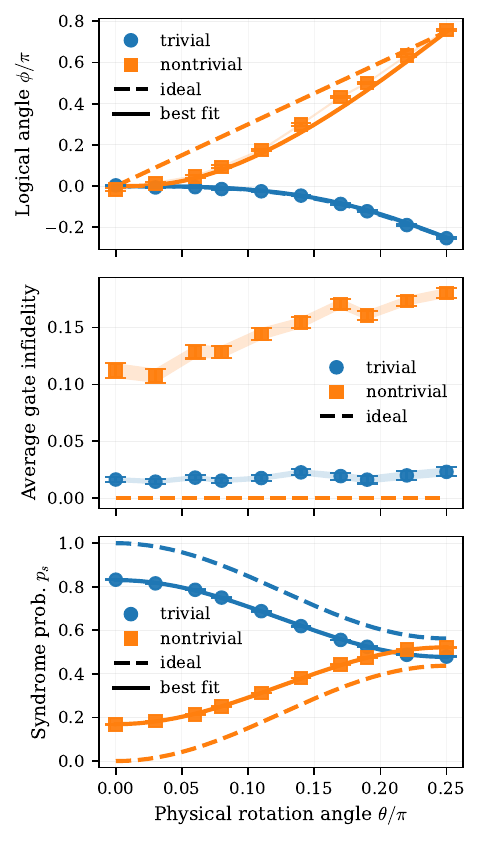}
      \put(-2,98){\colorbox{white}{\textbf{(a)}}}
      \put(-2,67){\colorbox{white}{\textbf{(b)}}}
      \put(-2,34){\colorbox{white}{\textbf{(c)}}}
    \end{overpic}
  \end{center}
  \caption{%
    One-round protocol tomography results.
    \textbf{(a)}
    Logical rotation angle $\phi_s$ versus physical rotation angle $\theta$ for
    the trivial (blue circles) and nontrivial (orange squares) syndrome groups.
    \textbf{(b)}
    Residual average gate infidelity after removing the best-fit logical
    rotation.
    In both cases the ideal average gate infidelity (dashed lines) is
    uniformly zero.
    \textbf{(c)}
    Syndrome probabilities $p_s(\theta)$.
    Points with error bars show the reconstructed data, shaded bands indicate
    $1\sigma$ uncertainties, dashed curves show the ideal $p=0$ prediction, and
    solid curves show the best-fit simplified dephasing model, obtained by
    fitting the syndrome probabilities with $p=(2.62 \pm 0.01)\%$.
  }%
  \label{fig:1rtomo}
\end{figure}

The extracted channel parameters are summarized in \cref{fig:1rtomo}.
Panel (a) shows the syndrome-conditioned logical rotation angle
\(\phi_s(\theta)\).
For the trivial syndrome branch, the observed rotation closely follows the
ideal coherent prediction over the full angular range, confirming that a single
round of the protocol implements a controllable logical \(Z\) rotation.
The nontrivial branch also exhibits a coherent rotation, but with larger
deviations from the ideal curve.
Its overall trend is still reasonably described by the best-fit simplified
dephasing-only model plotted as a solid line.

Panel (b) quantifies the residual noise after removing the best-fit
logical rotation in terms of the average gate infidelity of $\mathcal{E}'_s$.
Panel (c) shows the syndrome probabilities \(p_s(\theta)\),
which are well-described by the simplified dephasing model, whereas the
residual-noise data in panel (b) show that this model does not capture the full structure
of the effective logical channel, especially in the nontrivial branch.
Thus, the simplified model accurately captures the leading dependence of the
branch probabilities and coherent rotation angle, but it does not provide a
complete description of the residual logical noise.
The model parameter is fit only to the syndrome probabilities in panel (c),
and no additional fit is made to the residual infidelity in panel (b).

\begin{figure}[h]
  \begin{center}
    \includegraphics[width=0.95\columnwidth]{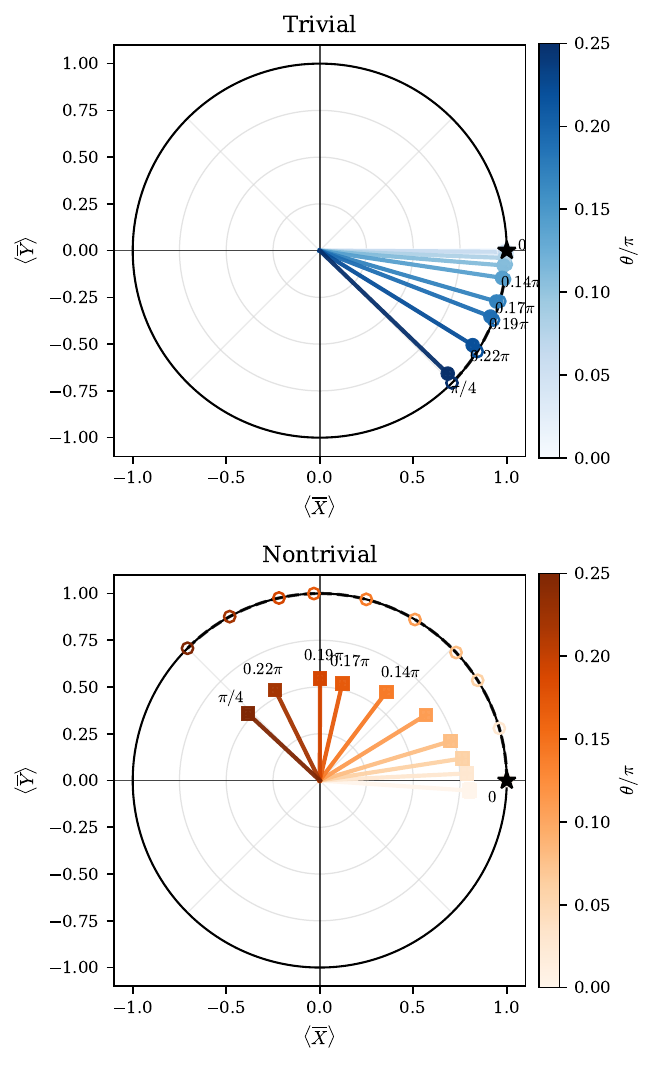}
  \end{center}
  \caption{%
    Logical output states projected onto the \(XY\) plane of the Bloch sphere for
    input \(\ket{\overline{+}}\) after one round of the protocol,
    marked as solid squares.
    \textbf{Top:} Trivial syndrome branch.
    \textbf{Bottom:} Aggregated nontrivial syndrome branch.
    For comparison, the ideal rotated states are marked as hollow circles of
    corresponding color.
  }%
  \label{fig:bloch1r}
\end{figure}

The larger residual infidelity of the nontrivial branch is consistent with the
fact that this branch corresponds to detected faults and is therefore more
sensitive to imperfections in state preparation, syndrome extraction, decoding,
and measurement, as well as physical Pauli \(X\) and \(Y\) errors, none of
which are included in the simplified dephasing-only model.
This behaviour is also visible in the Bloch-sphere representation shown in
\cref{fig:bloch1r}.
The trivial branch traces a cleaner rotation in the logical \(XY\) plane,
whereas the nontrivial branch shows substantially stronger contraction and
distortion.
This is consistent with the small code distance and with the role of the
nontrivial branch in the broader adaptive protocol, where highly noisy branches
would typically be disfavored or discarded.

\begin{figure}[h]
  \begin{center}
    \includegraphics[width=0.95\columnwidth]{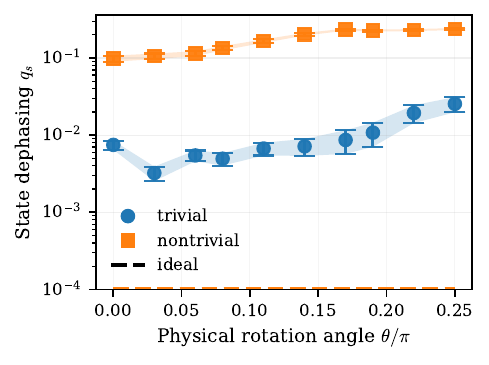}
  \end{center}
  \caption{%
    Logical dephasing $q_s(\theta)$ of the residual channel for variable
    physical rotation angle $\theta$ starting from the input state
    $\ket{\overline{+}}$ fitted from tomography data in the $\overline{X}$ and
    $\overline{Y}$ measurement bases.
  }%
  \label{fig:1rdephasing}
\end{figure}

Using only the input state $\ket{\overline{+}}$ to fit output $\overline{X}$
and $\overline{Y}$ basis tomography measurements to a dephasing and
$Z$-rotation channel gives the logical dephasing $q_s(\theta)$ of the residual
channel as shown in \ref{fig:1rdephasing}.

Taken together, these tomography measurements provide a fuller channel-level
characterization than Ramsey interferometry alone.
They confirm the syndrome-conditioned coherent logical rotation while also
revealing additional residual noise beyond the dephasing-only model.

\subsection{Logical tomography of the two-round protocol}

We next study how the logical channel composes over multiple rounds by
performing logical tomography on a two-round protocol.
In this experiment, we apply a transversal rotation by angle \(\theta\),
perform syndrome extraction and correction, then apply a second transversal
rotation by angle \(-\theta\) followed by a second round of syndrome
extraction.
In the ideal limit, the syndrome-conditioned logical rotation angles from the
two rounds should add.
In particular, if the same syndrome branch is obtained in both rounds, the
logical phases are expected to approximately cancel, so that the net logical
rotation is close to zero.
This can be seen in the dephasing-model where the logical rotation angle
$\phi_s(p, -\theta) = -\phi_s(p, \theta)$ has odd symmetry in $\theta$.

\begin{figure}[h]
  \input{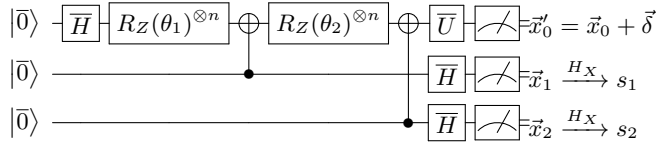}
  \caption{%
    Two-round logical state-tomography circuit.
    A logical input state \(\ket{\overline{+}}\) undergoes a round of the
    transversal rotation protocol with physical angle \(\theta\), followed by a
    second round with angle \(-\theta\).
    The output is conditioned on the pair of syndrome outcomes from the two
    rounds, allowing reconstruction of the effective logical channel in each
    syndrome pair.
    The correction is absorbed as a correction $\vec{\delta}$ for classical
    postprocessing.
  }%
  \label{fig:tworound}
\end{figure}

The two-round circuit is shown in \cref{fig:tworound}, with deferred correction
in classical postprocessing.
The IonQ Forte platform used does not support mid-circuit measurement and
feedforward,
so ancilla qubits cannot be reset and reused.
As each syndrome extraction round requires 16 fresh physical ancilla qubits,
the 36-qubit limit of the register would be insufficient for two rounds.
We thus restrict ourselves to doing only syndrome extraction of
$X$-type stabilizer generators, which requires only 8 extra physical qubits
each round.
Accordingly, in this experiment we restrict to logical state tomography with
input state \(\ket{\overline{+}}\), which is sufficient to probe the
syndrome-resolved rotation and dephasing of the effective logical channel in
the logical \(XY\) plane.
We group the data by the syndrome outcomes of the first and second rounds,
yielding four syndrome classes:
trivial-trivial, trivial-nontrivial, nontrivial-trivial, and
nontrivial-nontrivial.

\begin{figure}[h]
  \begin{center}
    \includegraphics[width=0.95\columnwidth]{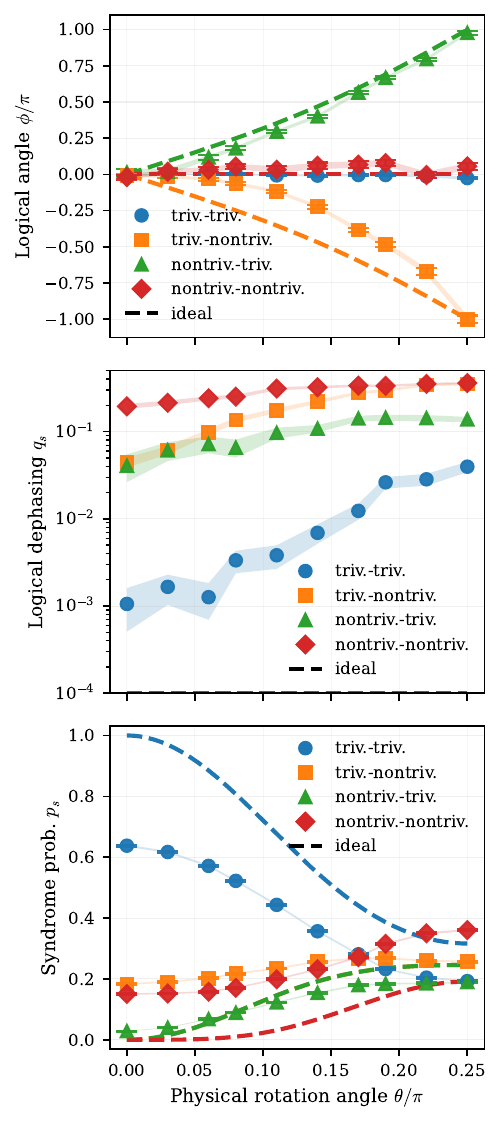}
  \end{center}
  \caption{%
    Two-round logical tomography results for input state \(\ket{\overline{+}}\).
    \textbf{Top:} Fitted logical rotation angle versus physical angle \(\theta\)
    for each syndrome pair.
    \textbf{Middle:} Corresponding fitted logical dephasing parameter \(q_s\).
    \textbf{Bottom:} Syndrome-pair probabilities versus \(\theta\), with ideal
    curves as dashed lines.
  }%
  \label{fig:2rtomo}
\end{figure}

The extracted two-round channel parameters are shown in \cref{fig:2rtomo},
where we assume the state takes the form
$\mathcal{D}_{q_s} \circ \mathcal{R}_Z(\phi_s)
(\ket{\overline{+}}\bra{\overline{+}})$.
The top panel plots the fitted logical rotation angle $\phi_s$ for each
syndrome class as a function of the applied physical angle \(\theta\), while
the middle panel shows the corresponding fitted logical dephasing parameter
\(q_s\).
The bottom panel shows the syndrome-pair probabilities, together with the ideal
curves.
Overall, the measured logical rotation angles are consistent with the expected
error model: the different syndrome pairs produce distinct effective logical
rotations, and branches with the same syndrome in both rounds show the
strongest cancellation.
Most notably, to within error bars,
the trivial-trivial branch remains close to zero logical
rotation over the full angular range and exhibits the lowest fitted dephasing,
indicating the intended cancellation of \(+\theta\) and \(-\theta\).

As expected the other syndrome pairs show larger residual logical rotations and
increased dephasing,
since a nontrivial syndrome in either round signals a fault,
explainable by multiple possible error mechanisms.
The trivial-nontrivial and nontrivial-trivial branches need not be equivalent:
they condition on a detected fault at different points in the circuit, so a
first-round fault can affect the state entering the second round whereas a
second-round fault cannot.
The observed order asymmetry is therefore plausible, but quantitatively
identifying its source would require a more detailed time-resolved noise model
and is left to future work.
We do not attempt to fit these non-trivial cases to the dephasing-only model
since it would be an oversimplification of the error mechanisms.

\begin{figure*}[h]
  \includegraphics[width=0.95\textwidth]{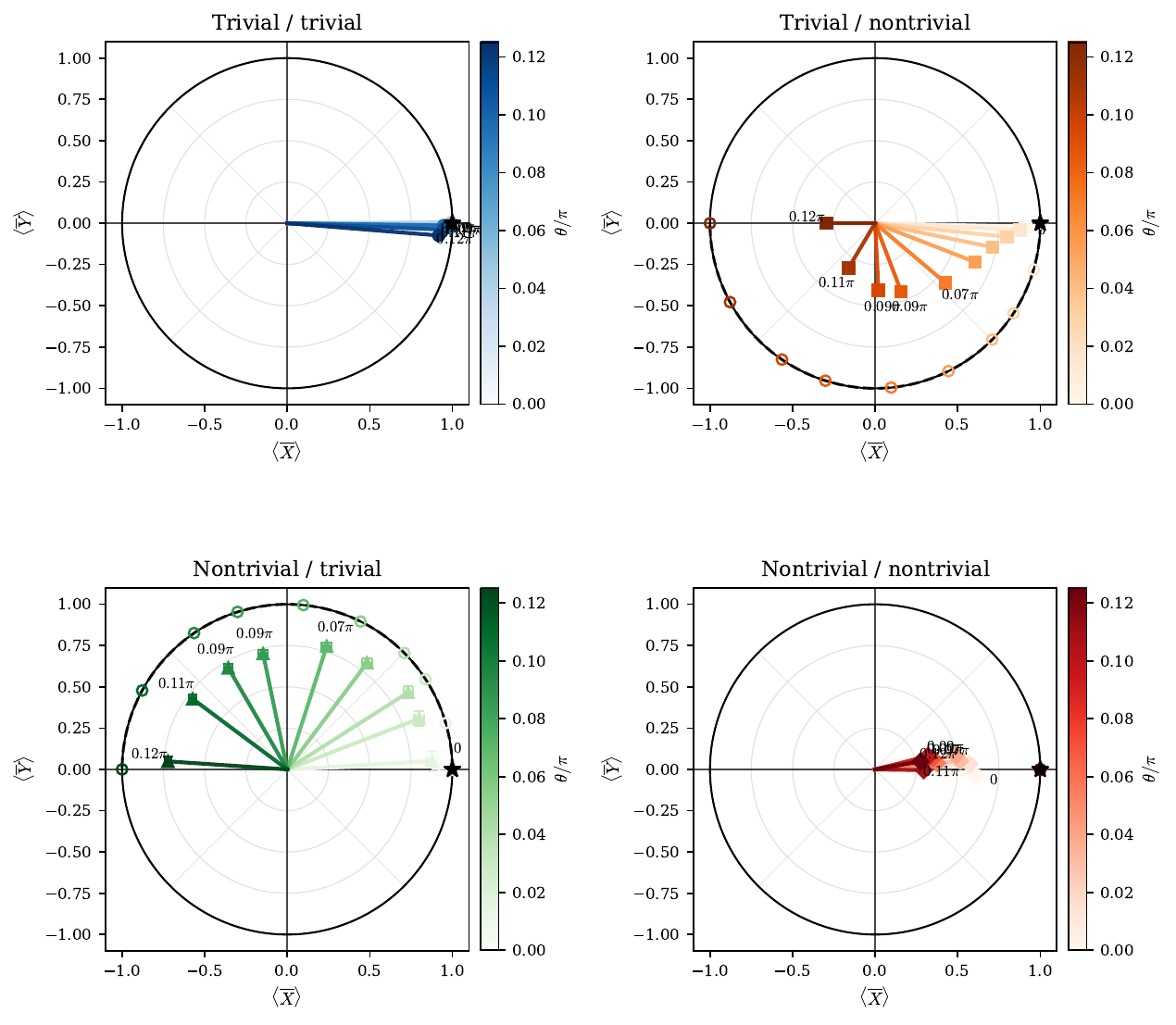}
  \caption{%
    Logical output states on the \(XY\) plane of the Bloch sphere for the
    two-round tomography experiment with input state
    \(\ket{\overline{+}}\), resolved by syndrome pair.
    Each square point corresponds to a distinct physical rotation angle \(\theta\), as
    labeled, with error bars indicating \(1\sigma\) uncertainties.
    Correspondngly colored hollow markers show the ideal noiseless rotated
    states for reference.
  }%
  \label{fig:bloch2r}
\end{figure*}

This behavior is visualized directly in the Bloch-plane representation shown in
\cref{fig:bloch2r}.
For an input state \(\ket{\overline{+}}\), the ideal two-round cancellation
would return the logical Bloch vector close to its starting point on the
positive \(X\) axis.
Instead, the measured outputs trace syndrome-dependent trajectories that both
rotate and contract.
The trivial-trivial branch remains clustered nearest to the initial state,
while the branches containing nontrivial syndromes show progressively larger
angular offsets and stronger contraction toward the origin.
Thus the two-round experiment reveals both approximate additivity of the
syndrome-conditioned logical angles and persistent accumulation of noise across
rounds.

Taken together, these results show that the logical action of the protocol
continues to be structured and interpretable beyond a single round.
Although the small code distance and hardware constraints prevent a fully
fault-tolerant multi-round adaptive implementation,
it shows logical angles compose reasonably while noisy syndromes can be
readily identified.

\section{Discussion}\label{sec:discussion}

We have analytically derived the expected channel parameters of the transversal
rotation protocol in the Steane code under a dephasing-only noise model.
We then demonstrated that transversal physical $Z$ rotations combined
with Steane error correction can realize continuous-angle logical $Z$
rotations in the $[[7,1,3]]$ Steane code, with the implemented logical
operation conditioned on the measured syndrome.
The one-round Ramsey and tomography data show that the leading coherent part of
the logical channel is well described by a syndrome-dependent logical rotation
$\phi_s$,
whose behavior is well-described by a simplified dephasing-only error model
that fits excellently with the observed syndrome probabilities.
The two-round experiment further shows that these syndrome-resolved logical
angles compose in a structured way across rounds:
when the two rounds yield the same syndrome class, the logical phases
approximately cancel for the $+\theta$ and $-\theta$ sequence, while branches
containing nontrivial syndromes exhibit larger residual rotations and stronger
dephasing.

Although this model captures the dominant dependence of the branch
probabilities and the logical rotation angle $\phi_s$, the residual
infidelity indicates additional error mechanisms beyond simple physical
dephasing.
These likely include imperfections in flagged state preparation,
syndrome extraction, coherent calibration errors,
measurement error, physical Pauli \(X\) and \(Y\) errors, and correlated faults introduced by the relatively
deep circuits required for logical state preparation, syndrome extraction, and
logical-basis measurement.
Because a nontrivial syndrome heralds a detected fault, these branches are
expected to be more sensitive to such imperfections, consistent with the larger
dephasing and distortion observed in the tomography.

An important practical limitation of the present experiment is that the logical
corrections were deferred to classical postprocessing rather than applied by
real-time feedforward.
Accordingly, our implementation should be viewed as a characterization of the
syndrome-resolved logical channels generated by the protocol, rather than a
fully adaptive fault-tolerant realization.
In addition, the 36-qubit hardware limit precluded full two-round Steane error
correction, so the two-round state tomography experiment was restricted to
$X$-syndrome extraction and logical state tomography for input
\(\ket{\overline{+}}\).
These constraints limit quantitative performance, but they do not obscure the
main qualitative observation that the protocol generates knowable,
continuously tunable logical phases that remain interpretable over multiple
rounds.

More broadly, these results provide an experimental proof of principle for
continuous logical control beyond the logical Clifford group using only
transversal coherent rotations, syndrome measurements, and classical decoding.
For larger-distance codes and hardware with higher-fidelity gates and
lower-latency feedforward, one may envision implementing the fully adaptive
multi-round protocol of \cite{huangRobustPhaseContinuous2026}, in which the
physical rotation angle in each round is chosen based on the previously
observed syndromes to approach the desired target logical rotation efficiently.
An important next step will therefore be to combine this protocol with improved
fault-tolerant state preparation and syndrome extraction on larger codes
supporting more than just one syndrome branch with low logical dephasing
so that the fault-tolerance of the adaptive continuous-angle logical protocol
can be tested at scale.

\begin{acknowledgements}
  We thank Milan Kornjača for insightful discussions.
  E.H. is supported by the Fulbright Future Scholarship.  This material is based upon work supported in part by the
Defense Advanced Research Projects Agency (DARPA)
under Agreement HR00112490357, the NSF QLCI award OMA2120757, and the NSF-funded NQVL:QSTD: Design: ORAQL.  Certain commercial equipment, instruments, or materials are identified in this paper in order to specify
the procedure adequately and do not reflect any endorsement by NIST.
\end{acknowledgements}

\bibliographystyle{apsrev4-2}
\bibliography{papers}

\appendix

\newpage
\end{document}